\documentclass[aps, prd, twocolumn, superscriptaddress, nofootinbib, showpacs,floatfix]{revtex4}
\usepackage[dvips]{epsfig}

\usepackage{natbib}

\def\lsim{\lower0.6ex\vbox{\hbox{$ \buildrel{\textstyle <}\over{\sim}\ $}}}
\def\gsim{\lower0.6ex\vbox{\hbox{$ \buildrel{\textstyle >}\over{\sim}\ $}}}

\def\beq{\begin{equation}}
\def\eeq{\end{equation}}

\def\beqa{\begin{eqnarray}}
\def\eeqa{\end{eqnarray}}

\def\bfig{\begin{figure}[ht] \begin{center}}
\def\bfigh{\begin{figure}[h!] \begin{center}}
\def\bfigb{\begin{figure}[hb!] \begin{center}}
\def\bfigt{\begin{figure}[t!] \begin{center}}
\def\bfight{\begin{figure}[ht!] \begin{center}}
\def\efig{\end{center} \end{figure}}

\def\btab{\begin{table*}[ht]}
\def\etab{\end{table*}}

\def\aap{Astron.\ \& Astrophys.}
\def\mnras{Mon. Not. Roy. Astron. Soc.}

\begin{document}

\title{Constraining Sterile Neutrino Warm Dark Matter with \it{Chandra}\rm~\bf{Observations of the Andromeda Galaxy}}

\author{Casey R. Watson}
\affiliation{Department of Physics and Astronomy,
Millikin University, Decatur, Illinois 62522}

\author{Zhiyuan Li}
\affiliation{Harvard-Smithsonian Center for Astrophysics,
Cambridge, MA 02138}

\author{Nicholas K. Polley}
\affiliation{Department of Physics and Astronomy,
Millikin University, Decatur, Illinois 62522}



\begin{abstract}

We use the \it{Chandra}\rm~unresolved X-ray emission spectrum
from a $12'-28'$ (2.8-6.4 kpc) annular region of the Andromeda galaxy
to constrain the radiative decay of sterile neutrino warm dark 
matter. By excising the most baryon-dominated, central 2.8 kpc of the galaxy,
we reduce the uncertainties in our estimate of the dark matter mass within the field of view
and improve the signal-to-noise ratio of prospective sterile neutrino decay signatures
relative to hot gas and unresolved stellar emission.
Our findings impose the most stringent limit on the sterile neutrino mass to date in 
the context of the Dodelson-Widrow model, $m_s < 2.2 \rm\ keV$ (95\% C.L.). Our results also
constrain alternative sterile neutrino production scenarios 
at very small active-sterile neutrino mixing angles.

\end{abstract}

\maketitle

\section{Introduction}

In nearly all extensions of the Standard Model, the
generation of neutrino masses leads to the
introduction of sterile neutrinos, e.g., \cite{Dinh:2006ia,Dong:2006mt}.
Sterile neutrinos may also play the role of 
the dark matter, e.g., \cite{Dodelson1994,Shi:1998km,AFP,AFT,Dolgov:2000ew,AbaSav,Taoso:2007qk,
Bazzocchi:2008fh, Boyarsky:2009ix, Wu:2009yr, Gelmini:2009xd, Bezrukov:2009th} and 
affect a host of interesting cosmological and astrophysical processes, including 
the production of baryon \cite{AS2005,ABS2005,Asaka:2006ek,Smith:2008ic} and lepton \cite{ABFW} 
asymmetries, Big Bang Nucleosynthesis \cite{Dolgov:2000jw,ABFW,Smith:2008ic}, the evolution of the
matter power spectrum \cite{Boyanovsky:2008nc,Boyanovsky:2010pw}, reionization \cite{ 
Barkana:2001gr,Yoshida:2003rm,Hansen:2003yj,Gelmini:2004ah,Biermann:2006bu,O'Shea:2006tp,Mapelli:2006ej, 
Ripamonti:2006gq,Stasielak:2006br}, 
neutrino oscillations \cite{Cirelli:2004cz,Smirnov:2006bu},
pulsar kicks \cite{Kusenko:1998bk,Fuller:2003gy,Barkovich:2004jp,Kusenko:2004mm,Fryer:2005sz,Kusenko:2008gh},
and supernovae \cite{Hidaka:2006sg,Hidaka:2007se,Raffelt:2011nc}. 
The degree to which sterile neutrinos participate in 
these processes is sensitive to the strength of their interactions with Standard Model 
particles and their abundance throughout cosmic history. The creation and exploration of
models and production scenarios 
that characterize these properties has been
\cite{Dodelson1994,Shi:1998km,AFP,AFT,Dolgov:2000ew}
and continues to be \cite{Aba05a,Aba05b,Shaposhnikov:2006xi,deGouvea:2006gz, Kusenko:2006rh,
Asaka:2006nq,Petraki:2007gq,Khalil:2008kp,
Kadota:2007mv,Lattanzi:2008ds,Kishimoto:2008ic,
Shaposhnikov:2008pf,Laine:2008pg,Jenkins:2008fx,Bazzocchi:2008fh, Boyarsky:2009ix,
Wu:2009yr, Gelmini:2009xd, Bezrukov:2009th} a very active area of research.
See Ref.~\cite{Kusenko:2009up} for a review of sterile neutrino properties.

In addition to their rich phenomenology, sterile neutrinos are also a readily testable 
dark matter candidate. To date, their properties have primarily been constrained in 
two ways: through X-ray searches for
their radiative decays and via observed cosmological small-scale structure, but the possibility
of detection through atomic ionization and nuclear spin flip signatures has also recently been discussed
\cite{Ando:2010ye}.
Although the two primary methods are generally considered separately, 
combined studies of both radiative and cosmological constraints have also been performed 
\cite{Palazzo:2007gz,Boyanovsky:2007ay,Bazzocchi:2008fh}.  

The radiative decays of predominantly sterile neutrino mass eigenstates to
predominantly active neutrino mass eigenstates and X-rays of energy $E_{\gamma, \rm
s} = m_s/2$ can be detected by existing X-ray satellites \cite{AFT}.
Current radiative decay limits are based on
observations of a long list of sources, including the Cosmic X-ray background
\cite{Boyarsky:2005us,Cumberbatch:2007qq}, nearby clusters such as Virgo \cite{Aba05b} and Coma
\cite{AbaSav,Boyarsky:2006zi}, more distant clusters like A520 and A1835
\cite{RiemerSorensen:2006pi} and the bullet cluster \cite{Boyarsky:2006kc}, nearby galaxies
like Andromeda \cite{Watson:2006, Boyarsky:2007ay}, M33 \cite{Borriello:2011un},
Milky Way satellite galaxies like the Large
Magellenic Cloud \cite{Boyarsky:2006fg}, Ursa Minor \cite{Boyarsky:2006ag, Loewenstein:2008yi},
Draco \cite{RiemerSorensen:2009jp}, Wilman I \cite{Loewenstein:2009cm}, and Seque I \cite{Mirabal:2010an},
as well as unresolved emission from the Milky Way itself
\cite{Boyarsky:2006ag,RiemerSorensen:2006fh,Abazajian:2006jc,Yuksel:2007xh,Boyarsky:2007ge}.
See Refs.~\cite{AbaSav,Watson:2006,Abazajian:2006jc} and references therein for a more detailed summary
and Refs.~\cite{Yuksel:2007dr,deVega:2009ku} for more general studies of radiative limits.

Although the original Dodelson-Widrow (DW) non-resonant sterile neutrino production scenario
\cite{Dodelson1994, AFT, Aba05b} has nearly been excluded by radiative constraints alone, e.g, $m_s < 3.5$ keV
(95\% C.L.) \cite{Watson:2006},
and can account for at most 70\% of the dark matter at 2$_{~}\sigma$ according to
Ref.~\cite{Palazzo:2007gz}, recent Suzaku observations of Wilman I revealed
a spectral feature at 2.5 keV that was consistent with the radiative decay of a 5 keV DW sterile neutrino
\cite{Loewenstein:2009cm}. These findings have been discussed further
in Refs.~\cite{Boyarsky:2010ci,Kusenko:2010sq}, and we will return to
them in Sections~\ref{Analysis}~and~\ref{conclusions} of this paper.

In addition to the
DW scenario, many other models have been proposed that predict the proper
relic dark matter density even at very small active-sterile neutrino mixing angles, e.g.,
\cite{Shi:1998km, AFP, AbaSav, Petraki:2007gq, Laine:2008pg}.  General phase space considerations
have also been used to constrain sterile neutrino production scenarios, e.g.,
\cite{Boyarsky:2008ju, Gorbunov:2008ka}.

As a complement to the upper bounds imposed by radiative decay
constraints, observations of the clustering of cosmological structures on the smallest scales can set
lower bounds on the mass of the dark matter particle.
This is because sufficiently light dark matter particles would have suppressed or
even erased these smallest structures by easily propagating (free-streaming) out
of the shallow gravitational potential wells in which they formed.

Current cosmological lower limits on $m_s$ are
based primarily on measurements of small scale clustering in the
CMB, SDSS galaxy, and Ly$\alpha$ forest flux power spectra and gravitational lensing
by the smallest structures. However, the results of N-body
simulations have also recently been used to set lower bounds on $m_s$ based on the requirement that
the number of simulated satellite galaxies equals or exceeds the number of observed Milky Way
satellites \cite{Polisensky:2010rw}. 

Of all the cosmological constraints, the Ly$\alpha$ limits are the most sensitive to the
difficult-to-characterize, non-linear growth of baryonic and dark matter structures as well as
the specifics of dark matter production, e.g., the momentum distribution of the
dark matter particles. There have been significant discrepancies between the lower
limits \cite{Aba05b,Seljak:2006qw,Viel:2006kd,Viel:2007mv,Boyarsky:2008xj,Boyarsky:2008mt}
obtained using different data sets and hydrodynamical simulations, as discussed in, e.g.,
\cite{AbaSav,Abazajian:2006jc}.
Ly$\alpha$ limits are less restrictive for
models in which sterile neutrinos behave more like cold dark matter, e.g., 
\cite{AbaSav,Asaka:2006ek,Petraki:2008ef}. 

Studies of gravitationally lensed galaxies and QSOs have provided comparable lower limits,
e.g., $m_s > 5.2$ keV \cite{Dalal:2002su} and $m_s > 
10$ keV \cite{Miranda:2007rb}, respectively, based on the image distortions caused by the smallest structures.
These lensing constraints should improve significantly as data from future submillilensing 
experiments become available \cite{Hisano:2006cj}.  Lower bounds based on comparisons of satellite galaxy
counts to N-body simulations,
e.g., $m_s > 13.3$ keV (in the DW scenario) \cite{Polisensky:2010rw}, are also similar to the most restrictive
Ly$\alpha$ constraints.

In this paper, we consider the radiative decay limits imposed by \it{Chandra}\rm~\cite{Weisskopf:2002}~observations 
of the Andromeda galaxy (M31).  We choose to focus on Andromeda because of 1)
its close proximity, 2) its substantial and well-studied dark matter distribution,
and 3) its intrinsically low level of X-ray emission. These properties tend to maximize the
prospective sterile neutrino decay signal (1 and 2) and minimize noise, i.e., astrophysical background (3).

The \it{Chandra}\rm~data set we use in this study also has several advantages over 
the \it{XMM-Newton}\rm~observations of Andromeda \cite{Shirey} that were used to constrain the properties of 
sterile neutrinos in Ref.~\cite{Watson:2006} (W06).
First, compared to \it{XMM-Newton}\rm~EPIC, the
\it{Chandra}\rm~ACIS detector has a significantly lower and much more stable
instrumental background, making it particularly well-suited for extracting low surface
brightness, extended X-ray emission.
Second, using \it{Chandra's}\rm~superior angular resolution, we are able
to remove the emission from a larger number of resolved, X-ray point sources than would be possible
with \it{XMM-Newton},\rm~thereby further reducing astrophysical background noise,
while excising very little dark matter from the small excluded regions.
Third, the field of view (FOV) associated with the \it{Chandra}\rm~spectrum
we use in this paper is over 25 times the area of the \it{XMM-Newton}\rm~FOV studied in W06
(a 12 arcminute to 28 arcminute annulus vs.~a circle of 5 arcminute radius) and
therefore probes a significantly larger
dark matter mass and prospective $\nu_s$ decay signal.
The dark matter mass within this FOV is not only significantly larger but 
also subject to less uncertainty both because we
exclude the central region of the galaxy, where the dark
matter density profile is least well constrained, and because we make use of an updated model of the
Andromeda dark matter distribution
based on more recent kinematic data and theoretical considerations.
The results of this study are also more accurate because we 
use the \it{Chandra}\rm~ACIS-I effective area, rather than an average flux-to-count ratio, to convert 
calculated sterile neutrino decay signals into spectral features. 
For Majorana sterile neutrinos, the resulting unresolved emission spectrum requires $m_s  < 2.2$ keV 
(95\% C.L.) to avoid more than doubling the observed signal in bins of energy $E \ge$ 1.1 keV. 
Although the mass-mixing exclusion region we generate with this new data set is not vastly more
restrictive than the exclusion region presented in W06 (see Fig. 4 below), it is more robust for the reasons cited above,
and as we argue throughout the paper, it will be difficult to improve upon the constraints we set here with
the current generation of X-ray detectors.

In Sec.~\ref{Model}, we discuss the basics of the DW scenario as well as some more recent 
sterile neutrino models.
In Sec.~\ref{M31_prop}, we discuss the \it{Chandra}\rm~observations of Andromeda,
how we generate the unresolved X-ray spectrum we use for our constraints,
and how we calculate the dark matter mass contained within the \it{Chandra}\rm~FOV.
In Sec.~\ref{Analysis}, we present our analysis of the unresolved spectrum, discuss the 
limits we are able to impose on the sterile neutrino mass and mixing,
and compare our results to those of previous studies.
In Sec.~\ref{conclusions} we summarize our findings and discuss the outlook
for future sterile neutrino constraints.
Throughout the paper, we assume a flat cosmology 
with $\Omega_{\rm baryon}=0.04$, $\Omega_{\rm WDM}= \Omega_{\rm s}= 0.24$,
$\Omega_{\Lambda}=0.72$, and $h = H_0/100~ \hbox{km s}^{-1} ~\rm  Mpc^{-1} = 0.72$.

\section{The Sterile Neutrino Warm Dark Matter Model}
\label{Model}

The radiative decay rate for Majorana sterile neutrinos is \cite{PW,Barger}
\beq
\Gamma_s \simeq 1.36\times 10^{-32} \rm{s}^{-1} 
\left(\frac{\sin ^{2}2\theta}{10^{-10}}\right)
\left(\frac{m_s}{\rm~keV}\right)^5.
\eeq
The line flux at $ E_{\gamma ,\rm s} = m_s/2$ resulting from the decay of
the fraction of sterile neutrinos that are within the detector field of view (FOV),
$N^{\rm FOV}_s = (M^{\rm FOV}_{\rm DM}/m_s )$ in a halo at a distance $D$
is given by \cite{AFT,Aba05a,Aba05b}
\beqa
\Phi_{\rm x,s}(\sin ^{2}2\theta) = \frac{E_{\gamma ,\rm s}N^{\rm FOV}_{\rm s}\Gamma_s} {4\pi D^2} \simeq
    1.0 \times 10^{-17} \rm{erg~cm}^{-2}\rm{s}^{-1} \nonumber \\
    \times \left(\frac{M^{\rm FOV}_{\rm DM}}{10^{11}M_\odot }\right)
    \left(\frac{D}{\rm Mpc}\right)^{-2}  
    \left(\frac{\sin ^{2}2\theta}{10^{-10}}\right) 
    \left(\frac{m_s}{\rm~keV}\right)^5,
\label{Phixs}
\eeqa
where $\sin ^{2}2\theta$ characterizes the active-sterile neutrino mixing.
Dirac sterile neutrinos would produce only half the flux given by Eqn.~(\ref{Phixs}) \cite{PW,Barger}.
To facilitate comparisons between our work and other results, we will adopt Majorana sterile neutrinos,
which have been more commonly assumed in recent studies,
e.g., \cite{Boyarsky:2007ge,Abazajian:2006jc,Boyarsky:2007ay}\footnote{The decay rate
equation given in Ref.~\cite{Abazajian:2006jc}
is for Majorana rather than Dirac sterile neutrinos.}.

For a QCD phase-transition
temperature of $T_{\rm QCD} = 170$ MeV 
and a lepton asymmetry of $L \simeq n_{\rm baryon}/n_\gamma \simeq 10^{-10}$,
the sterile neutrino density-production 
relationship for this model is \cite{Aba05a}
\beq
m_s = 55.5\rm~{keV}\left(\frac{sin^{2}2\theta}{10^{-10}}\right)^{-0.615}
\left(\frac{\Omega_{\rm s}}{0.24}\right)^{0.5}. 
\label{Omega_sin2th}
\eeq
Combining Eqns.~(\ref{Phixs}) and (\ref{Omega_sin2th}) yields an expression for the line flux 
that is independent of 
the mixing angle:
\beqa
\Phi_{\rm x,s}(\Omega_{\rm s}) \simeq 7.0 \times 10^{-15} \rm{erg~cm}^{-2}\rm{s}^{-1} \nonumber \\
    \times \left(\frac{M^{\rm FOV}_{\rm DM}}{10^{11}M_\odot }\right)
    \left(\frac{D}{\rm Mpc}\right)^{-2}  
    \left(\frac{\Omega_{\rm s}}{0.24}\right)^{0.813}
    \left(\frac{m_s}{\rm~keV}\right)^{3.374}.
\label{Phixs_nomix}
\eeqa

We note that the density-production relationship presented in Ref.~\cite{Asaka:2006nq}
agrees with Eqn.~(\ref{Omega_sin2th}) for sterile neutrino masses ranging from
1 keV $\lsim m_{s~} \lsim 10$ keV. The sterile neutrino mass limits we determine in 
Sec.~\ref{Analysis} are therefore valid for the relationships in both
Refs.~\cite{Aba05a} and \cite{Asaka:2006nq}. It would also
be straightforward to re-evaluate our limits on $m_s$ assuming different 
density-production relationships, such as the three resonant production models 
associated with large $L$ \cite{AbaSav} or the Shi-Fuller model \cite{Shi:1998km} as 
calculated in Ref.~\cite{Laine:2008pg}, all of which are shown in Fig.~\ref{ms_mix} below.

\begin{figure}
\includegraphics[height = .35\textheight, width = .5\textwidth]{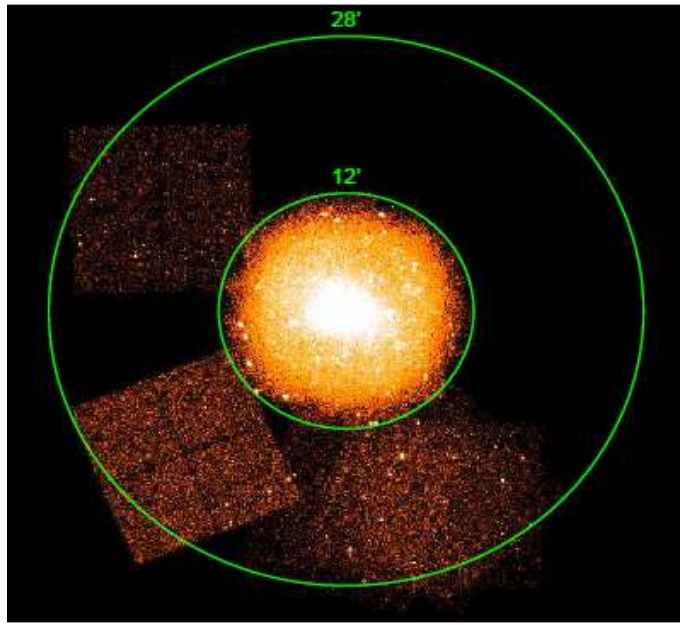}
\caption {Here we show the raw counts associated with the 7 \it{Chandra}\rm~ACIS-I
exposure regions (some of which are overlapping) within a $12'-28'$ annulus about the center
of the Andromeda galaxy. Variations in the brightness among these 7 regions are mainly
due to differences in exposure times (which vary from 5 ks to 20 ks) rather than intrinsic variation
in the emission (see text).
We have also included an ACIS-I image of the central $12'$ to illustrate the extent of the
X-ray emission from hot gas and point sources that we have eliminated by
excising the bulge of the galaxy from our FOV.
\label{ACIS_FOV}}
\end{figure}
   
\section{Properties of Andromeda}
\label{M31_prop}

\subsection{X-ray Data}

To detect or place the most restrictive constraints on radiatively decaying dark matter, 
it is critical to minimize the X-ray background from baryonic sources,
such as diffuse hot gas and X-ray binaries. In M31, both
the diffuse hot gas \cite{Li:2007ud,Li:2009mh}
and
X-ray binaries \cite{Voss:2006az}
exhibit a relatively steep radial distribution toward the galactic
center, suggesting that we can optimize the dark matter signal to
baryonic noise ratio by avoiding the inner regions of the galaxy.
Although most existing {\it{Chandra}} observations of M31 have been aimed toward
the inner bulge, we did find seven observations that were aimed toward the outer
bulge/disk regions and were therefore more well-suited for our purposes.
All seven observations used the Advanced CCD Imaging Spectrometer
(ACIS) I-array as the primary detector, with a $17'$x$17'$ field of view (FOV).
The positions of these observations relative to an annulus of
inner radius 12 arcminutes and outer radius 28 arcminutes ($12' - 28' \simeq 2.8 - 6.4$ kpc at
$D_{\rm M31} \simeq$ $0.78 \pm 0.02$ Mpc \cite{Stanek:1998cu,Jerjen:2003ws})
are shown in Figure~\ref{ACIS_FOV}.  
Five of the seven observations (ObsIDs 1576, 1584, 2899, 2901 and
2902) have effective exposure times of $\sim$5 ks, while the other two
observations have longer exposure times of 8 ks (ObsID 7067) and 20 ks
(ObsID 11256), respectively.  Each $17'$x$17'$ ACIS-I
region covers 11.4\% of the $12'-28'$ annulus, and
the non-overlapping parts of these 7 pointings collectively
cover $\sim 35\%$.  We address sampling
bias later in this section, pointing out that the co-added spectrum from
the selected observations provides, if anything, an overestimate of
the X-ray emission from the entire $12'-28'$ annulus.

We obtained and re-processed the archival data using CIAO version
4.2 and the corresponding calibration files. We followed the
standard procedure of data reduction, e.g., as described in Ref.~\cite{Li:2007ud}.
Briefly, for each observation, we (i) filtered time intervals of
high particle background, (ii) detected and removed discrete sources,
(iii) extracted a spectrum for the unresolved X-ray emission in the
0.5-8 keV range and generated corresponding instrumental response (rmf
and arf) files, and (iv) extracted a fiducial instrumental background
spectrum, using the ``stow background data'' that has been calibrated with
the 10-12 keV count rate.
In a final step, we co-added the
seven spectra and generated exposure-weighted response files, using
the {\sl FTOOL addspec}. Such a procedure is appropriate, since the spectra
were extracted from nearly identical detector regions.

Since the effective area $A_{\rm eff}(E)$ of ACIS is known to decrease gradually with time,
it is important to note that these seven observations were taken over a time span of 9 years.
By using a weighted $A_{\rm eff}(E)$ associated with the co-added spectrum in our analysis below, we are
able to gauge the mean sensitivity of the detector over this time interval.
Additionally, we note that the degradation of the detector mainly
affects energies below 1 keV, and our data set has very little constraining power
at these low energies, as we will show.

It is also noteworthy that the observed flux varies modestly
among the seven observations, within a factor of 1.5 of the
mean values associated with the co-added spectrum. This is a natural
result, since the observations sampled different regions of the
outer bulge and the disk. We note, however, that the sampled regions mostly covered
the southeastern half of the disk -- where the
unresolved X-ray emission is higher than that from the
northeaster side, primarily due to the tilt of M31's
galactic plane \cite{Li:2007ud}. This sampling bias means that the co-added spectrum we use
represents an upper limit on the mean flux
within the $12'-28'$ annulus - precisely what we need
to set a robust upper limit on the mass of the sterile neutrino.
Finally, we note that a correction factor of $(0.114)^{-1}$ has been applied to the
data shown in Fig.~\ref{Galspec} to account for the fact that
the co-added spectrum from the 7 archived observations represents
11.4\% of the projected area of the $12'-28'$ annulus.

\subsection{Dark Matter Enclosed within the \it{Chandra}\rm~\bf{Field of View}}
\label{DMest}

The dark matter halo of M31 has been studied extensively.
Klypin, Zhao and Somerville (KZS) \cite{KZS}, for instance, used a variety of kinematic data to determine its dark 
matter density profile, $\rho_{\rm DM, M31}$.
Using more recent data and a more accurate baryonic mass profile, Seigar, Barth and Bullock
(SBB) \cite{Seigar:2006ia} have updated the profile found by KZS.
When we integrate the SBB model of $\rho_{\rm DM, M31}$ over 
the volume, $V_{\rm FOV, M31}$, associated
with the projection of the Chandra FOV along the line of sight,
we find that the $12' - 28'$ annulus contains:
\beqa
\Sigma^{\rm FOV}_{\rm DM,M31} \simeq (0.8 \pm 0.08) \times 10^{11} M_\odot \rm{Mpc}^{-2}; \nonumber \\
M^{\rm FOV}_{\rm DM,M31} \simeq (0.49 \pm 0.05) \times 10^{11} M_\odot,
\label{M_DM}
\eeqa  
where
\beq
\Sigma_{\rm FOV} = \int \frac{\rho_{\rm DM}(|\vec{r}-\vec{D}|)dV_{\rm FOV}}{ r^{2}}
\label{Sigma}
\eeq
is the dark matter column density within $V_{\rm FOV}$,
and the dark matter mass within $V_{\rm FOV}$ is defined by $M^{\rm FOV}_{\rm DM} = D^{2}\Sigma^{\rm FOV}_{\rm DM}$.
(See W06 for further details).

In addition to being based on more contemporary data and theoretical considerations, the SBB value
for $\Sigma^{\rm FOV}_{\rm DM,M31}$ is also more conservative:
less than 85\% of the KZS estimate for this annulus. We further note that for our FOV, CDM models predict
more conservative $\Sigma^{\rm FOV}_{\rm DM,M31}$ values than cored, WDM models. Because WDM
models are less centrally concentrated than CDM models, they are constrained to predict higher densities at larger radii
in order to reach agreement with the total mass of a given halo. For instance, within the radial limits
of our FOV (2.8-6.4 kpc), the dark matter density given by the Burkert profile for M31
\cite{Burkert:1995yz} with the commonly adopted parameter values (as in, e.g., Ref.~\cite{Boyarsky:2007ay})
is $17\% - 29\%$ higher than that of the KZS profile
and $18\% - 41\%$ higher than that of the SBB profile. 
Therefore, despite the fact we are constraining a WDM candidate in this paper,
we chose the SBB profile over a WDM profile in the interest of providing a
conservative dark matter mass estimate.

Because of \it{Chandra's}\rm~excellent angular resolution,
we lose less than 5\% of $\Sigma^{\rm FOV}_{\rm DM,M31}$
due to the excision of point sources. However,
to be conservative, we use only $0.95_{~} \Sigma^{\rm FOV}_{\rm DM,M31}$ for $\Sigma^{\rm FOV}_{\rm DM}$
in Eqns.~(\ref{dN_dE_dt_Om}) and (\ref{dN_dE_dt_sin}) to determine
the prospective sterile neutrino signals in the figures below,
and we ignore the contribution from the fraction of the Milky Way dark matter halo within the FOV.  

\section{Constraining Sterile Neutrino Decays}
\label{Analysis}

\begin{figure}
\includegraphics[height = .35\textheight, width = .5\textwidth]{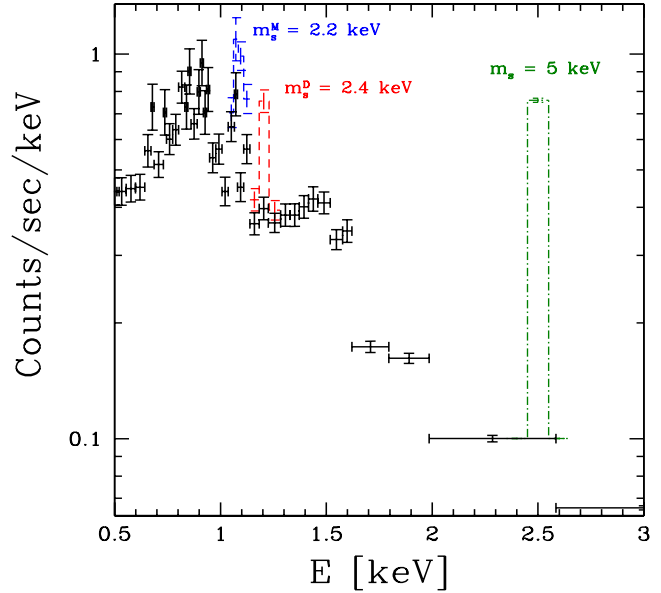}
\caption {Here we show the \it{Chandra}\rm~unresolved X-ray spectrum emitted from a $12'-28'$ annular
region about the center of the Andromeda galaxy (solid) and the 
first statistically significant $\nu_s$ decay 
peaks (dashed) at $E_{\gamma, \rm s} = m_{\rm s, lim}/2$ = 1.1 keV (blue) and 1.2 keV (red),
which exclude $m^{\rm M}_{\rm s} > 2.2$ keV (Majorana) and $m^{\rm 
D}_{\rm s} > 2.4$ keV (Dirac), respectively (95\% C.L.; Eqn.~\ref{m_s_limit_criterion}).
As a gauge of the statistical significance of our limits, we have included the ($1_{~}\sigma$) 
Poisson error bars on the unresolved emission spectrum data points. As a point of comparison to the
potential detection discussed in Ref.~\cite{Loewenstein:2009cm}, we also show
the decay signature that would be produced if the dark matter within the $12'-28'$ region of Andromeda were composed
of 5 keV sterile neutrinos (green, dot-dashed), a possibility that is strongly excluded by the data.
\label{Galspec}}
\end{figure}

In Fig.~\ref{Galspec}, we show the unresolved emission spectrum from the inner $12'-28'$ of 
Andromeda.
To calculate the $\nu_s$ decay signals, we 
assume that sterile neutrinos comprise all of the dark matter, i.e.,
$\Omega_{\rm s} = \Omega_{\rm DM} = 0.24$,
and evaluate Eqn.~(\ref{Phixs_nomix}) based on 95\% of the $\Sigma^{\rm FOV}_{\rm DM,M31}$ value
given in Eqn.~(\ref{M_DM}).
We then convert 
the sterile neutrino decay fluxes to the same units as those of the 
measured spectrum (Counts/sec/keV) as follows:
\beqa
\frac{dN_{\gamma, \rm s}}{dE_{\gamma, \rm s} dt} \left(\Omega_{\rm s}\right) = 
\left(\frac{\Phi_{\rm x,s}(\Omega_{\rm s})}{E_{\gamma, \rm s}}\right)
\left(\frac{A_{\rm eff} (E_{\gamma, \rm s})}{\Delta E}\right) 
\nonumber \\
%
%
= 6.7 \times 10^{-2}~\rm{Counts/sec/keV}
\left(\frac{\rm A_{\rm eff} (E_{\gamma, \rm s})}{100~\rm{cm}^2}\right) \nonumber \\
\times \left(\frac{\Sigma^{\rm FOV}_{\rm DM}}{10^{11}M_\odot \rm Mpc^{-2}}\right)
\left(\frac{\Omega_{\rm s}}{0.24}\right)^{0.813}
\left(\frac{m_s}{\rm~keV}\right)^{1.374},
\label{dN_dE_dt_Om}
\eeqa
where 1 erg/$E_{\gamma, \rm s} = 1.6 \times 10^{9}(E_{\gamma, \rm s}/\rm{keV})^{-1}$ gives the number 
of Counts associated with 1 erg at energy $E_{\gamma, \rm s}$ and
$A_{\rm eff} (E_{\gamma, \rm s})$ and $\Delta E$ are the
effective area\footnote{http://cxc.harvard.edu/proposer/POG/html/ACIS.html} and spectral energy resolution of the 
\it{Chandra}\rm~ACIS-I detector, respectively. To realistically simulate detected sterile neutrino decay 
``line'' fluxes in Fig.~\ref{Galspec}, we use a Gaussian centered at $E_{\gamma , s} = m_s/2$ 
with a FWHM of $\Delta E = E_{\gamma , s}/15$, a conservative estimate of the energy 
resolution of ACIS-I in imaging mode\footnote{http://heasarc.nasa.gov/docs/cxo/cxo.html}. 

\begin{table}[t!]
\caption{The Table shows the distance to Andromeda, the angular range, $\Delta \theta_{\rm FOV}$, of the annular field 
of view (FOV) probed by the \it{Chandra}\rm~observations, the dark matter 
column density, $\Sigma^{\rm FOV}_{\rm DM}$, enclosed within the FOV (Eqn.~\ref{M_DM}), and the (95\% C.L.) 
upper bounds on $m_s$ for Dirac and Majorana sterile neutrinos.}
\begin{ruledtabular}
\begin{tabular}{cc}
Galaxy Name&                                  Andromeda (M31)\\ \hline 
Distance (Mpc)&                           $0.78 \pm 0.02$\\ \hline
$\Delta \theta_{\rm FOV}$ (arcminutes)&          $12' -  28'$\\ \hline
$\Sigma^{\rm FOV}_{\rm DM}/10^{11} M_\odot \rm{Mpc}^{-2}$&   $0.8 \pm 0.08$\\ \hline
\textbf{$m^{\rm D}_{\rm s}$} (keV) (95\% C.L.)&         \textbf{2.4} (Dirac)\\ \hline
\textbf{$m^{\rm M}_{\rm s}$} (keV) (95\% C.L.)&         \textbf{2.2} (Majorana)\\ \hline
\end{tabular}
\end{ruledtabular}
\end{table}

\subsection{DW Sterile Neutrino Mass Limits}
\label{mass_limits}

To determine the mass limit imposed by the unresolved emission spectrum of Andromeda, we find the first 
bin of energy $E_{\gamma , s} = m_s/2$ for which the sterile neutrino decay signal 
(Eqn.~\ref{dN_dE_dt_Om}) at least doubles the amplitude of the measured spectrum in that bin, 
$\Delta {\cal F}$:    
\beq
\frac{dN_{\gamma, \rm s}}{dE_{\gamma, \rm s} dt}\left(\Omega_{\rm s}\right) \geq \Delta {\cal 
F}. 
\label{m_s_limit_criterion}
\eeq
The resulting limits: $m^{\rm M}_{\rm s} > 2.2$ keV (Majorana) and $m^{\rm
D}_{\rm s} > 2.4$ keV (Dirac), which are shown in Fig. 2 and Table I, are much more significant than the 95\% C.L.
defined by the ($1_{~}\sigma$) Poisson error bars on the measured points.
(For this data set, Eqn.~(\ref{m_s_limit_criterion}) is equivalent to $\gsim 4~\sigma$).
However, because we do not have a 
precise understanding of the features of the unresolved emission spectrum, which originate from some 
combination of hot gas, unresolved stellar sources, sky and detector backgrounds, and possibly sterile neutrino 
decay, we choose a limiting criterion that would remain robust even to 100\% level 
fluctuations in the data. (See W06 \cite{Watson:2006} 
for a more detailed discussion).

\begin{figure}
\includegraphics[height = .35\textheight, width = .5\textwidth]{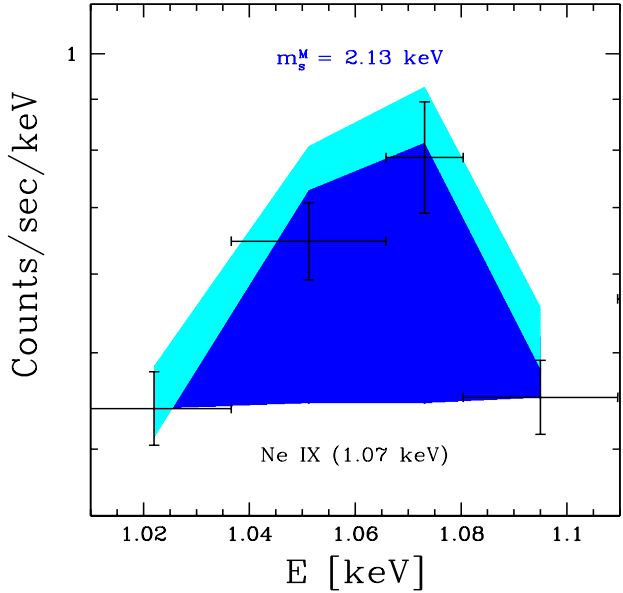}
\caption {Here, to illustrate how difficult it is to distinguish between atomic and anomalous
line features with current detectors, we show the statistical consistency of
the 1.07 keV Ne IX emission peak (\it{Chandra}\rm~data points in black)
and the decay signature of a 2.13 keV Majorana
sterile neutrino (blue, shaded region).
The light blue band shows the 1 $\sigma$ uncertainty range associated
with the \it{Chandra}\rm~data. 
\label{detection_features}}
\end{figure}

\subsection{Examining Possible DW Sterile Neutrino Decay Signatures}
\label{detection_tests}

There are a large number of atomic emission lines at energies
just below our exclusion limit, i.e., $E_{\gamma, \rm s} \lsim 1$ keV.
Differentiating between these features and anomalous lines will be difficult
if not impossible given the spectral resolution of current detectors.
The diminishing amplitude of sterile neutrino decay signatures at lower
energies, $\frac{dN}{dE dt} \propto  E_{\gamma, \rm s}^{1.374}$ (Eqn. \ref{dN_dE_dt_Om})
further exacerbates this problem.  Even with a data set such as ours, with a favorably high
ratio of FOV dark matter to astrophysical background, prospective sterile neutrino signals
become comparable to or dwarfed by atomic line features at energies just below 1.1 keV.

To illustrate this point, we examined whether or not any of the atomic emission lines below 1.1 keV in Fig.~2 were
statistically consistent with sterile neutrino decay.  We found that this was the case for the Ne IX peak at
1.07 keV. In particular, when we assumed that the Ne IX peak was a sterile neutrino decay line superposed
on a continuum background characterized by the mean flux values of the
bins to the left and right of the peak energy, we found that it was best fit by the decay signature of a 2.13 keV Majorana sterile neutrino (see Fig.~\ref{detection_features}).

This exercise underscores both the spectral resolution and amplitude issues.
If Andromeda's dark matter halo were actually composed of 2.13 keV sterile neutrinos,
they would produce a decay signature at 1.065 keV, but this is clearly indistinguishable
from the 1.07 keV Ne IX feature with the detectors aboard \it{Chandra}.\rm~If the sterile neutrinos
were of even lower mass, the amplitude of their decay signatures would be dwarfed among the
thicket of sub-keV atomic lines, and higher spectral resolution would become indispensible if
we were to have any hope of finding them.  One of the few options available in the absence of
such advances is to consider anomalous line ratios, as in Ref.~\cite{Prokhorov:2010us}, but
doing so requires extremely precise knowledge of chemical abundances, plasma temperature,
etc. of the region(s) of the target galaxy being examined. To make progress without such
complications, particularly at $E_{\gamma, \rm s} \lsim 1$ keV, a new generation of much
higher spectral resolution detectors is required, as we discuss further in the Sec.~\ref{conclusions}.

\begin{figure}
\includegraphics[height = .4\textheight, width = .5\textwidth]{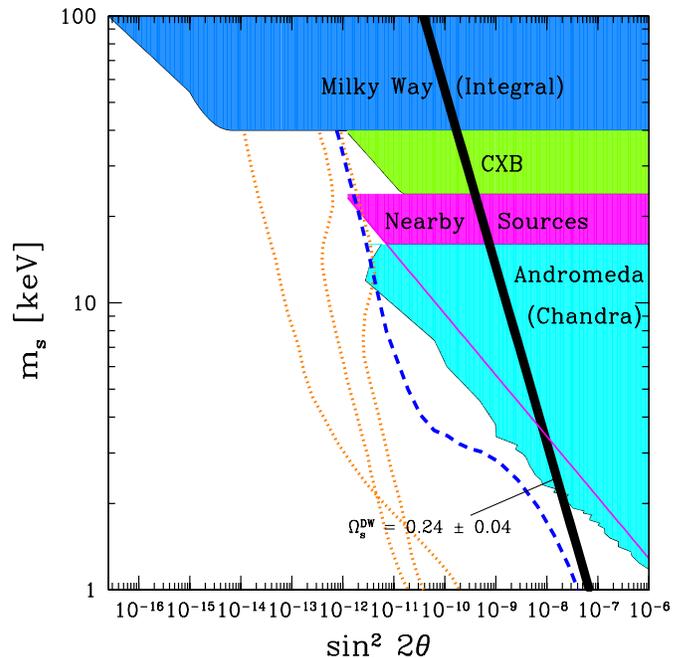}
\caption {Here we present constraints on $m_s$ as a function of 
mixing angle, sin$^2 2\theta$, assuming that all dark matter is comprised of sterile 
neutrinos ($\Omega_{\rm s} = 0.24$). 
For $L \simeq 10^{-10}$, the thick, solid line corresponds to $\Omega^{\rm DW}_{\rm s}= 0.24\pm 0.04$ in 
the Dodelson-Widrow (DW) scenario (Eqn.~\ref{Omega_sin2th}), while the region to the 
right corresponds to $\Omega^{\rm DW}_{\rm s} > 0.28$. Three density-production relationships 
associated with $\Omega_{\rm s}= 0.3$ and (left to right) $L =$ 0.1, 0.01, and 0.003
are also shown (dotted) \cite{AbaSav}, as is the Shi-Fuller density-production
relationship computed in Ref.~\cite{Laine:2008pg} (dashed). 
The three previous radiative decay upper 
limits (all 95\% C.L.) are based on \it{Integral}\rm~measurements of the unresolved X-ray 
emission from the Milky Way halo \cite{Yuksel:2007xh,Boyarsky:2007ge}, HEAO-1 and
\it{XMM-Newton}\rm~observations of the Cosmic X-ray Background (CXB)\cite{Boyarsky:2005us}, and the most stringent
constraints \cite{Watson:2006} from the many limits imposed by nearby galaxies and clusters 
\cite{Aba05b,AbaSav,Boyarsky:2006zi,Watson:2006,Boyarsky:2006fg,Boyarsky:2006ag,Abazajian:2006jc}.
The magenta line shows the recalculated boundary of this exclusion region for
Majorana sterile neutrinos (see text), allowing for direct comparison between our results and those of
Ref. \cite{Watson:2006}.
The most restrictive radiative decay limits, from the present work (also 95\% C.L.), are based on 
\it{Chandra}\rm~observations of the Andromeda galaxy.
\label{ms_mix}}
\end{figure}

\subsection{Exclusion Regions in the Mass-Mixing Plane}
\label{exclusion}

To determine the region of the $m_{s~} -$~sin$^{2}2\theta$ (mass-mixing) plane (Fig.~\ref{ms_mix}) 
that is excluded by the unresolved X-ray spectrum of Andromeda, we convert 
Eqn.~(\ref{Phixs}) to Counts/sec/keV:
\beqa
\frac{dN_{\gamma, \rm s}}{dE_{\gamma, \rm s} dt} \left(\sin ^{2}2\theta\right) =
\left(\frac{\Phi_{\rm x,s}(\sin ^{2}2\theta)}{E_{\gamma, \rm s}}\right)
\left(\frac{A_{\rm eff} (E_{\gamma, \rm s})}{\Delta E}\right) 
\nonumber \\
%
%
= 9.8 \times 10^{-5}~\rm{Counts/sec/keV}
\left(\frac{\rm A_{\rm eff} (E_{\gamma, \rm s})}{100~\rm{cm}^2}\right)\nonumber \\
\times \left(\frac{\Sigma^{\rm FOV}_{\rm DM}}{10^{11}M_\odot \rm Mpc^{-2}}\right)
\left(\frac{\sin ^{2}2\theta}{10^{-10}}\right)
\left(\frac{m_s}{\rm~keV}\right)^{3},
\label{dN_dE_dt_sin}
\eeqa
and adopt the analog of Eqn.~(\ref{m_s_limit_criterion}) as our exclusion criterion:
\beq
\frac{dN_{\gamma, \rm s}}{dE_{\gamma, \rm s} dt}\left(\sin ^{2}2\theta \right) \geq \Delta 
{\cal F}. 
\label{mass_mixing_exclusion}
\eeq
Just as we found two mass limits, we also derived two exclusion regions for Dirac and Majorana 
sterile neutrinos. The most restrictive region, which was determined by comparing the
\it{Chandra}\rm~unresolved X-ray 
spectrum to the Majorana sterile neutrino decay flux, is shown in Fig.~\ref{ms_mix}.  The ``indentation'' of our
exclusion region at the highest masses comes about because the effective area of the ACIS-I detector falls
even more steeply than the spectral data at the highest photon energies.
As discussed in the conclusion, a new instrument
with a much larger effective area and superior spectral energy resolution will be required to dramatically improve
upon the radiative decay constraints presented here.

In addition to 
our new Andromeda bounds, 
(the distinct parts of) three previously determined radiative decay exclusion regions are also shown in 
Fig.~\ref{ms_mix}.  
The upper exclusion region is based on \it{Integral}\rm~measurements of the unresolved X-ray 
emission from the Milky Way halo \cite{Yuksel:2007xh}. We note that these results were
corroborated by a very similar later study \cite{Boyarsky:2007ge}.
The second exclusion region was derived by analyzing 
HEAO-1 and \it{XMM-Newton}\rm~observations of the Cosmic X-ray Background (CXB)\cite{Boyarsky:2005us}.
The third region represents the most stringent constraints \cite{Watson:2006} (W06)
from the many limits imposed by nearby galaxies and clusters
\cite{Aba05b,AbaSav,Boyarsky:2006zi,Watson:2006,Boyarsky:2006fg,Boyarsky:2006ag,Abazajian:2006jc}.
We note that if we were to recalculate the exclusion region derived in W06 \cite{Watson:2006} for Majorana rather than Dirac
sterile neutrinos, the boundary of the exclusion region would remain the same (the magenta line shown in Fig. 4).
This is because we would not only need to multiply $\Phi_{\rm x,s}$ by a factor of 2, we would also have
to change the overly optimistic estimate of the spectral energy resolution used in that paper ($\Delta E = E/30$)
to the more realistic value used in this analysis ($\Delta E = E/15$).
Since $\frac{dN_{\gamma, \rm s}}{dE_{\gamma, \rm s} dt} \propto \frac{\Phi_{\rm x,s}}{\Delta E}$ (Eqn. 9),
the factors of 2 offset.

Boyarsky et al. \cite{Boyarsky:2007ay} also conducted a very thorough analysis of an annular
(\it{XMM-Newton}\rm-observed) region of Andromeda ($5' - 13'$).
Unfortunately, because of \it{XMM-Newton's}\rm~poorer 
spatial resolution, they were forced to excise almost one fourth of the field of view ($\sim 23$\%) to 
remove the emission from point sources, thereby sacrificing potentially signal-producing dark 
matter to reduce the astrophysical background.  As a result, the unresolved emission spectrum
they generated probed
a much smaller dark matter mass than the \it{Chandra}\rm~unresolved X-ray spectrum we consider here,
and their limits are correspondingly less restrictive ($m_s < 4$ keV).

\section{Conclusions}
\label{conclusions}

In this paper, we used the \it{Chandra}\rm~unresolved X-ray spectrum of the Andromeda galaxy 
(M31) to improve the radiative decay constraints on sterile neutrino warm 
dark matter.
Assuming either the model described in Refs.~\cite{AFP,AFT,Aba05a,Aba05b} or 
Ref.~\cite{Asaka:2006nq}, our analysis requires $m_s < 2.4 \rm\ keV$ (95$\%$ C.L.), in the 
least restrictive case (Dirac sterile neutrinos).  Because 
many recent papers, e.g., \cite{Boyarsky:2007ge,Abazajian:2006jc,Boyarsky:2007ay} have assumed 
Majorana sterile neutrinos, we quote the Majorana limit:
\beq
m_s < 2.2 \rm~keV ~~~(\rm Majorana;~95\% ~\rm C.L.)
\label{M31_12_28_Majorana}
\eeq
as our lower bound for the DW scenario, so that comparisons to other studies can be made on equal 
footing.

Most of the currently available sterile neutrino mass-mixing 
parameter space remains open only for scenarios in which $\Omega_{\rm s} \sim 0.3$ can be 
generated at very small mixing. Sterile neutrinos that are, e.g., resonantly produced in the
presence of a large lepton asymmetry ($L \gg 10^{-10}$), e.g., \cite{Shi:1998km,AFP,AbaSav,Laine:2008pg}
or created via Higgs decays \cite{Petraki:2007gq} are still viable. However, the combination of our X-ray
constraints with those from nearby sources
are also able to partially restrict the $L=0.003$ resonant production
model \cite{AbaSav} and the Shi-Fuller model \cite{Laine:2008pg} for sterile neutrino masses in the
10 to 24 keV range (Fig. 4), and the \it{Integral}\rm~measurements of the Milky Way halo
\cite{Yuksel:2007xh,Boyarsky:2007ge} exclude
all the alternative production scenarios shown in Fig.~\ref{ms_mix} above $m_s = 40$ keV.

Even without cosmological small scale structure bounds, the DW scenario remains
viable only between the Tremaine-Gunn bound \cite{TG:1979} and our limit from this work,
0.4 keV $< m_s <$ 2.2 keV, which interestingly falls within the range of dark matter particle
masses that best explains the core of the Fornax Dwarf Spheroidal galaxy \cite{Strigari:2006ue}.
This result underscores the need to continue to carefully and independently pursue all possible constraints
on sterile neutrino properties. However, as noted above,
because of the decreasing sterile neutrino signal at lower $m_s$ values
and the large number of atomic emission features at energies $\lsim 1$ keV,
it will be difficult to improve upon the 2.2 keV limit we have presented here with existing
X-ray detectors.
One of the most promising routes toward improved radiative constraints
includes the use
of much higher spectral resolution instruments than those currently available, as discussed, e.g.,
in Refs.~\cite{Abazajian:2006jc,Boyarsky:2006hr,Abazajian:2009hx,Herder:2009im}.
The International X-ray Observatory
(\it{IXO}\footnote{\rm http://ixo.gsfc.nasa.gov}\rm ),~for instance, which is scheduled for launch in 2021,
will have a FOV comparable to \it{Chandra}\rm~detectors ($\sim$~18' for photon energies ranging from 0.1-15 keV),
but $\sim$~100 times their effective area, $\sim$~10 times their spectral resolution, and $\sim$~10 times lower
instrumental background.  Based on these specifications, a megasecond observation of Andromeda with IXO will have
the capacity to fully test the 4 alternative sterile neutrino production scenarios shown in Fig.~\ref{ms_mix}
over the entire range of mass-mixing parameter space for which they remain viable \cite{Abazajian:2009hx}.


\acknowledgments
We thank M. Garcia, S. Murray, and Q.D. Wang, the principal investigators responsible for conducting
the Chandra observations of M31 used in this work. We also thank John Beacom and Hasan Y{\"u}ksel
for extensive and very helpful discussions.
CRW and NP acknowledge support from the Millikin University 
Department of Physics and Astronomy. ZL acknowledges support from SAO grant GO0-11098B.

\end{document}